\documentclass[10pt]{wiley2sp}
\usepackage{url}
\usepackage{helvet}
\usepackage{epsfig}
\usepackage{times}
\usepackage{amsmath}
\usepackage{graphicx}%
\usepackage{subfigure}%
\def\sectcounterend{.}
\def\figurename{Figure}

\def\pages{}
\def\thevol{}
\def\thenumber{}
\def\theyear{}
\def\thedoi{}

\begin{document}

\titlefigure[clip,width=.75\columnwidth]{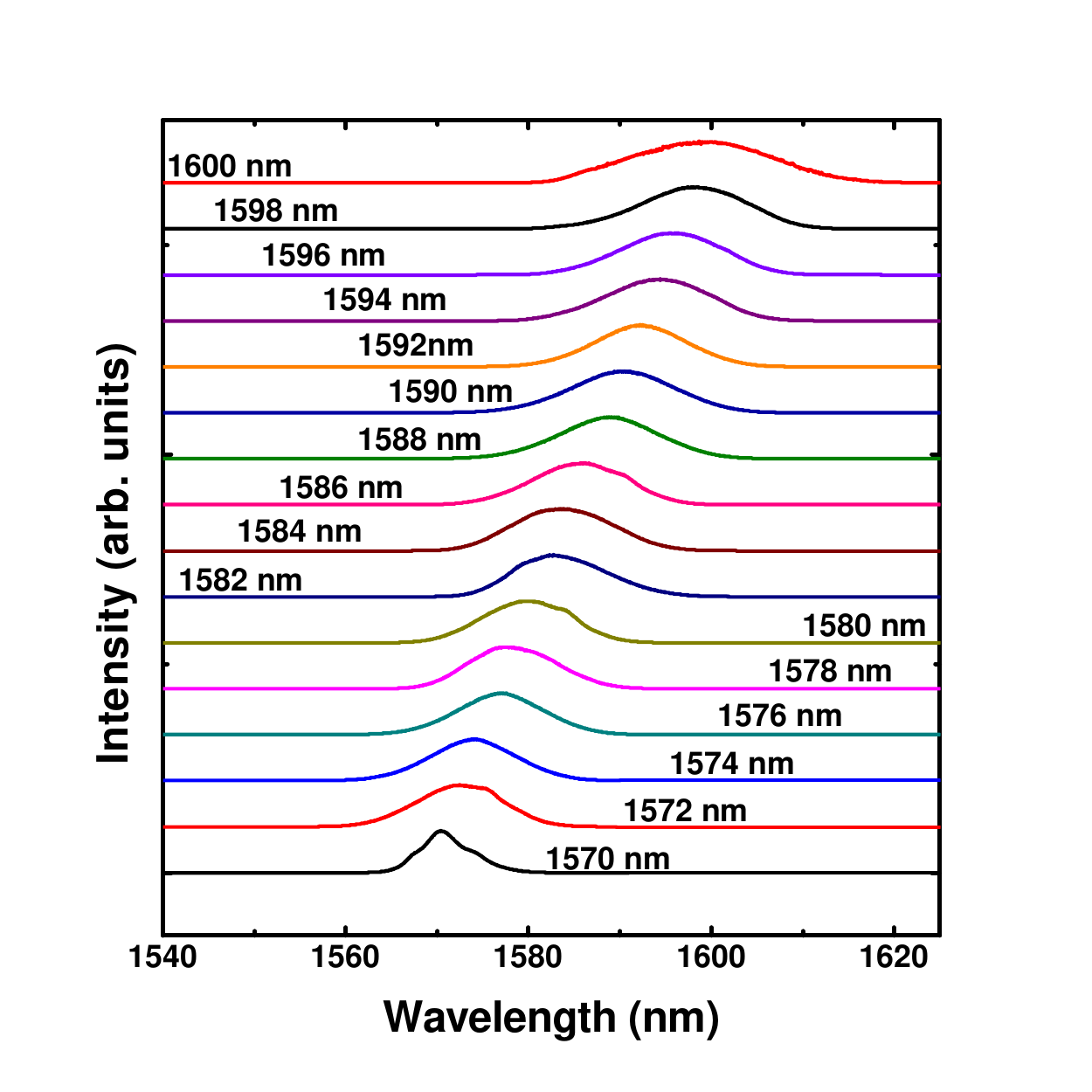}

\abstract{ Soliton operation and soliton wavelength tuning of erbium-doped fiber lasers mode locked with atomic layer graphene was experimentally investigated under various cavity dispersion conditions. It was shown that not only wide range soliton wavelength tuning but also soltion pulse width variation could be obtained in the fiber lasers. Our results show that the graphene mode locked erbium-doped fiber lasers provide a compact, user friendly and low cost wavelength tunable ultrahsort pulse source. }

\titlefigurecaption{Graphene mode locked wideband tunable output spectra from 1570 nm to 1600 nm}

\title{Compact graphene mode-locked wavelength-tunable erbium-doped fiber lasers: from all anomalous dispersion to all normal dispersion}

\author{H. Zhang,\inst{1} D. Y. Tang,\inst{1*} L. M. Zhao,\inst{1} Q. L. Bao,\inst{2} K. P. Loh,\inst{2} B. Lin,\inst{1}and S. C. Tjin\inst{1}}
\institute{1.School of Electrical and Electronic Engineering, Nanyang Technological University, Singapore 639798\\
2. Department of Chemistry, National University of Singapore, Singapore 117543}
 \mail{edytang@ntu.edu.sg}
 \received{}


\titlerunning{Compact graphene mode-locked wavelength-tunable erbium-doped fiber lasers}
\authorrunning{H. Zhang, D. Y. Tang, et al.}
\published{}

\maketitle
\sloppy

\section{Introduction}
Since its first isolation in 2004 \cite{ref1}, graphene has become one of the most shining materials and is now at the hub of scientific research.
Graphene is a single two dimensional atomic layer of carbon atoms forming a honeycomb crystal lattice.
It is a semiconductor that has a zero-bandgap and a linear energy dispersion relation for both electrons and holes \cite{ref1}.
Due to its novel quantum and transportation properties, grapheen has been intensively exploited for the high speed electronic devices applications \cite{ref2}.
However, graphene also possesses excellent photonic properties. It has been shown that graphene absorbs a constant fraction
of 2.3\% over a broad wavelength range from visible to the infrared, despite of the fact that it has only single atomic layer.
 This is an unexpectedly large absorption, considering that a 10 nm thick GaAs quantum well only absorbs about 1\% light with the near-gap frequency. Recently, the ultrafast carrier dynamics of graphene was experimentally investigated \cite{ref3}. It was shown that the carrier relaxation processes in graphene have ultrafast time constants. In view of the unique broad band absorption property of graphene, Xia et al. have firstly fabricated an ultra-fast graphene based photo-detector with ultrafast response \cite{ref4}; In previous papers we have also experimentally demonstrated that the wideband absorption of graphene could become saturated under intensive illumination \cite{ref5,ref6,ref7,ref8,ref9}. Taking advantage of the saturable absorption property of graphene, we have further demonstrated mode-locking of an erbium-doped fiber laser and a diode-pumped Nd:YAG ceramic laser, operating at the wavelength of 1.56  m and 1064 nm, respectively, with few layer graphene \cite{ref9}. We fabricated graphene-polymer nano-composite membrane to achieve soliton mode locking with per-pulse energy up to 3 nJ in anomalous dispersion cavity \cite{ref8} and used atomic layer graphene to obtain large energy mode locking with single pulse energy as high as 7.3 nJ \cite{ref7}.  However, wavelength-tunable graphene based mode locking in various dispersion regimes have not been reported. Although we had indeed experimentally studied whether semiconductor saturable absorber mirrors (SESAMs) could show the ability of wide-band tunable mode locking  in reference \cite{ref13}, unfortunately, we only observed very limited tuning regime, which can be traced back to the saturable absorption principle of SESAMs.  In contrast with the nowadays dominant SESAMs whose operation bandwidth are severely limited by the energy band-gap, graphene could be operated as a wideband saturable absorber with bandwidth broader than any other known material because of its unique electronic band-gap structure together with the central role of Pauli blocking in bleaching the light absorption. Our experiments clearly show that this advantage could render graphene rich applications in wavelength-tunable mode locked lasers.

 Wide range wavelength tunable mode locked lasers have widespread applications in laser spectroscopy, bio-imaging, and scientific researches. Although wavelength-tunable mode-locked lasers can be realized with the active mode locking technique, which is generally done by inserting an electro-optic modulator in the laser cavity, due to the bandwidth limitation of the modulator, the mode locked pulses thus obtained have broad pulse width. In addition, actively mode locked lasers are bulky and expensive. It is attractive if user-friendly, compact, low cost wavelength tunable mode locked lasers could be developed. Passively mode-locked soliton fiber lasers have the advantages of compact and low cost. It could also emit mode locked pulses with excellent pulse stability, low time jittering, ultra-short pulse width. Although passively mode locked fiber lasers have been extensively investigated in the past, recently it has also been shown that with the dissipative soliton formation technique, large energy mode locked pulses could be formed in the lasers \cite{ref10}, it still remains as a challenge to simultaneously achieve large energy ultrashort pulse generation and broadband wavelength tuning in the lasers. We note that L. E. Nelson et al. had demonstrated a wide range wavelength tunable erbium-doped fiber laser mode locked with the nonlinear polarization rotation technique (NPR). It was shown that through inserting a quartz birefringence plate in the cavity and changing the orientation of the plate, wide range wavelength tuning of the formed solitons could be obtained \cite{ref11}.  However, in their laser due to the intrinsic peak clamping effect of the NPR technique, the solitons formed in their laser have only small pulse energy. Recently, using the mixed diameter single wall carbon nanotubes as a saturable absorber, F. Wang et al. demonstrated a wide range wavelength tunable soliton erbium-doped fiber laser \cite{ref12}. Although SWCNTs with uniformly distributed diameters and chiralities can have broadband saturable absorption, an intrinsic drawback of the saturable absorber is that it has large non-saturable losses.

 Graphene has intrinsically broadband saturable absorption. In this letter, we show experimentally that one can exploit the broadband saturable absorption of graphene to construct wide range wavelength tunable mode locked fiber lasers. Graphene mode locked erbium-doped fiber lasers with various cavity dispersions were experimentally constructed, whose operation and wavelength tuning properties were experimentally investigated. We show that either the wavelength tunable transform-limited pulses or the large frequency chirped pulses could be generated.

\section{Experimental setups}

It is well-known that the erbium-doped fibers have a broad gain bandwidth of about 30nm in the 1550nm wavelength band \cite{ref11}.
Making use of the broad gain bandwidth of the erbium-doped fibers,
we have experimentally constructed three different cavity dispersion erbium-doped fiber lasers,
with the cavity dispersion varies from the all anomalous dispersion, to the near zero net cavity dispersion, and the all normal cavity dispersion.
We would like to compare the mode locking operation of the lasers and their wavelength tuning performance.
 Fig. 1 shows a schematic of the fiber laser configuration.
All the fiber lasers have a ring cavity, which comprises the erbium-doped fiber (EDF) as the gain medium and either the standard single mode fiber (SMF) or the dispersion
compensation fiber (DCF) to form other parts of the cavity. A polarization independent isolator was used to force the unidirectional operation of the ring cavity.
An intra-cavity polarization controller (PC) was used to adjust the cavity birefringence, and a 10\% fiber coupler was used to output the laser emission. The laser was pumped by a high power Fiber Raman Laser of wavelength 1480 nm. The few-layer graphene was used as the saturable absorber to mode lock the laser. The graphene saturable absorber was inserted in the cavity through transferring a piece of free standing few-layer graphene film onto the end facet of a fiber pigtail via the Van der Walls force. Details on the graphene or graphene-polymer nanocomposite thin-film preparation and characterization were reported in [5].
At first the soliton operation and wavelength tuning of an all-anomalous dispersion cavity erbium-doped fiber laser was investigated. In this case the 6.5 m EDF with a group velocity dispersion (GVD) parameter of 10 (ps/nm)/km, and a total length of 9.0 m SMF with GVD parameter of 18 (ps/nm)/km were used to construct the cavity, and the total cavity dispersion is estimated at $\sim$-0.293 ps$^2$. To change the cavity dispersion, we then replaced the anomalous dispersion EDF with a piece of 5 m normal dispersion EDF with GVD parameter of -32 (ps/nm)/km). Through varying the length of the SMF, we have adjusted the net cavity dispersion of the fiber laser. With a SMF length of 20.5 m, we estimate that the total net cavity dispersion is $\sim$-0.0026 ps$^2$. Finally, we have replaces the SMF in the cavity with the dispersion compensation fiber (DCF) with GVD parameter of -2 (ps/nm)/km and constructed an all-normal dispersion cavity erbium-doped fiber laser. With a DCF length of 9 m, the total dispersion of the all-normal cavity is about $\sim$ 0.231ps$^2$.
To achieve wavelength tuning of a mode locked laser, conventionally a tunable bandpass filter is inserted in the cavity to limit the effective laser gain bandwidth \cite{ref12}.  In our experiments in order to simplify the laser configuration, we have made use of the intrinsic artificial birefringence filter effect of the lasers \cite{ref13}. To this end we intentionally introduced strong cavity birefringence into the laser cavities through over bending the fibers looped in polarization controller.  In this way, the artificial birefringence filter effect of the cavity becomes no longer ignorable. Consequently, through changing the orientation of the intra cavity PC, the effective gain peak of the laser is shifted, and eventually the wavelength of the mode locked pulses is tuned.

\begin{figure}[t]
\vbox{\hbox
to\hsize{\psfig{figure=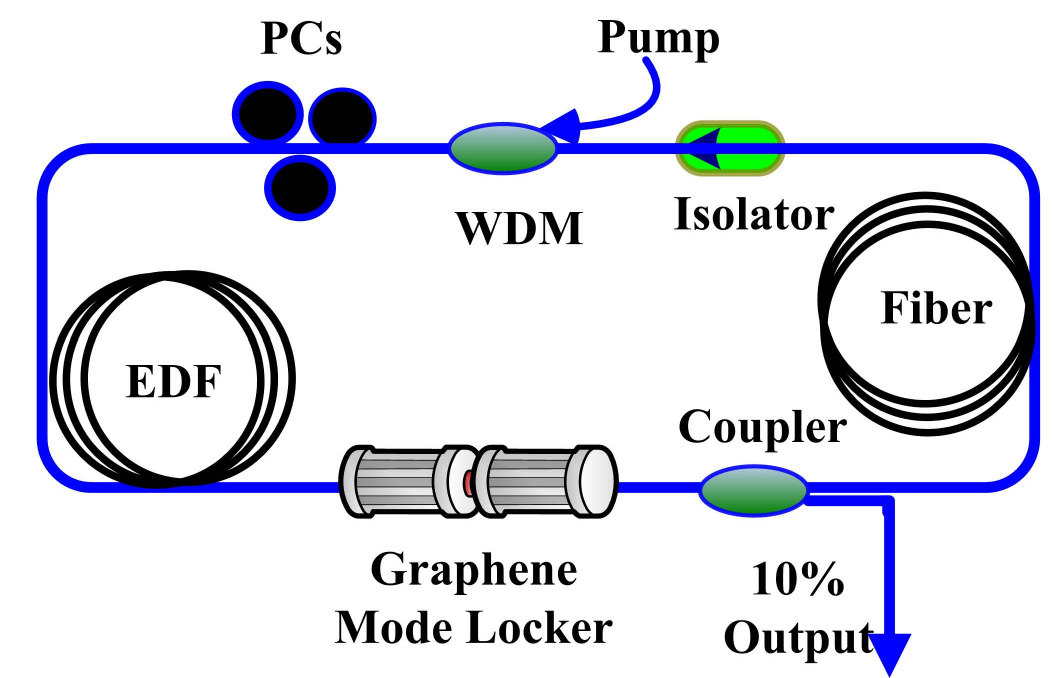,width=\hsize,clip=}\hfill}}
\caption{Schematic of the graphene mode locking fiber laser within different dispersion regimes. WDM denotes: wavelength division multiplexer, EDF denotes erbium-doped fiber, PC denotes polarization controller.}
\end{figure}

\section{Experimental results}
Operation of erbium-doped fiber lasers passively mode locked by either conventional NPR technique or SESAM in various cavity dispersion regimes were intensively investigated previously \cite{ref14,ref15,ref16,ref17,ref18,ref19,ref25,ref26,ref27,ref21}. It is well-known that in the anomalous cavity dispersion regime, the nonlinear Schrodinger equation type of soliton will be formed in the fiber laser, due to the natural balance between the anomalous cavity dispersion and fiber nonlinear optical Kerr effect, while in the large normal cavity dispersion regime, a dissipative soliton whose dynamics is described by the extended Ginzburg-Landau equation will be formed \cite{ref20}, and the formation of the soliton is a result of the mutual nonlinear interaction among the normal cavity dispersion, fiber Kerr nonlinearity, and the effective laser gain bandwidth filtering \cite{ref21}, moreover, in the regime of near zero cavity dispersion, the so called dispersion-managed solitons will be formed, which have a characteristic Gaussian pulse profile and power spectrum \cite{ref24}.

\subsection{Pure anomalous dispersion cavity}

Under graphene mode locking, we found that similar features of the soliton operation of the erbium-doped fiber lasers could be experimentally observed. Fig. 2a shows the mode locked solitons spectra and the soliton wavelength tuning obtained in the all-anomalous dispersion cavity fiber laser. The observed soliton spectra exhibit clear NLSE soliton features, characterized by the clear appearance of the Kelly sidebands. The laser has a mode locking threshold of about 100 mw. Once the soliton operation state is obtained, adjusting the orientation of the polarization controller, the wavelength of the solitons could be continuously tuned from 1560 nm to 1570 nm. The 3-dB bandwidth of the spectra is about 4.1 nm, which was almost unchanged as the central wavelength of the soliton was tuned. Fig. 2b shows the measured autocorrelation trace of the solitons. It has a FWHM width of about 1.0 ps. Assume a sech$^2$ pulse profile, the actual pulse width is about 655 fs. The time-bandwidth product (TBP) of the pulses is 0.335, indicating that they are transform-limited.

\begin{figure}[htbp]
  \centering
  \subfigure[]{
    \label{fig:subfig:a} 
    \includegraphics[width=8cm]{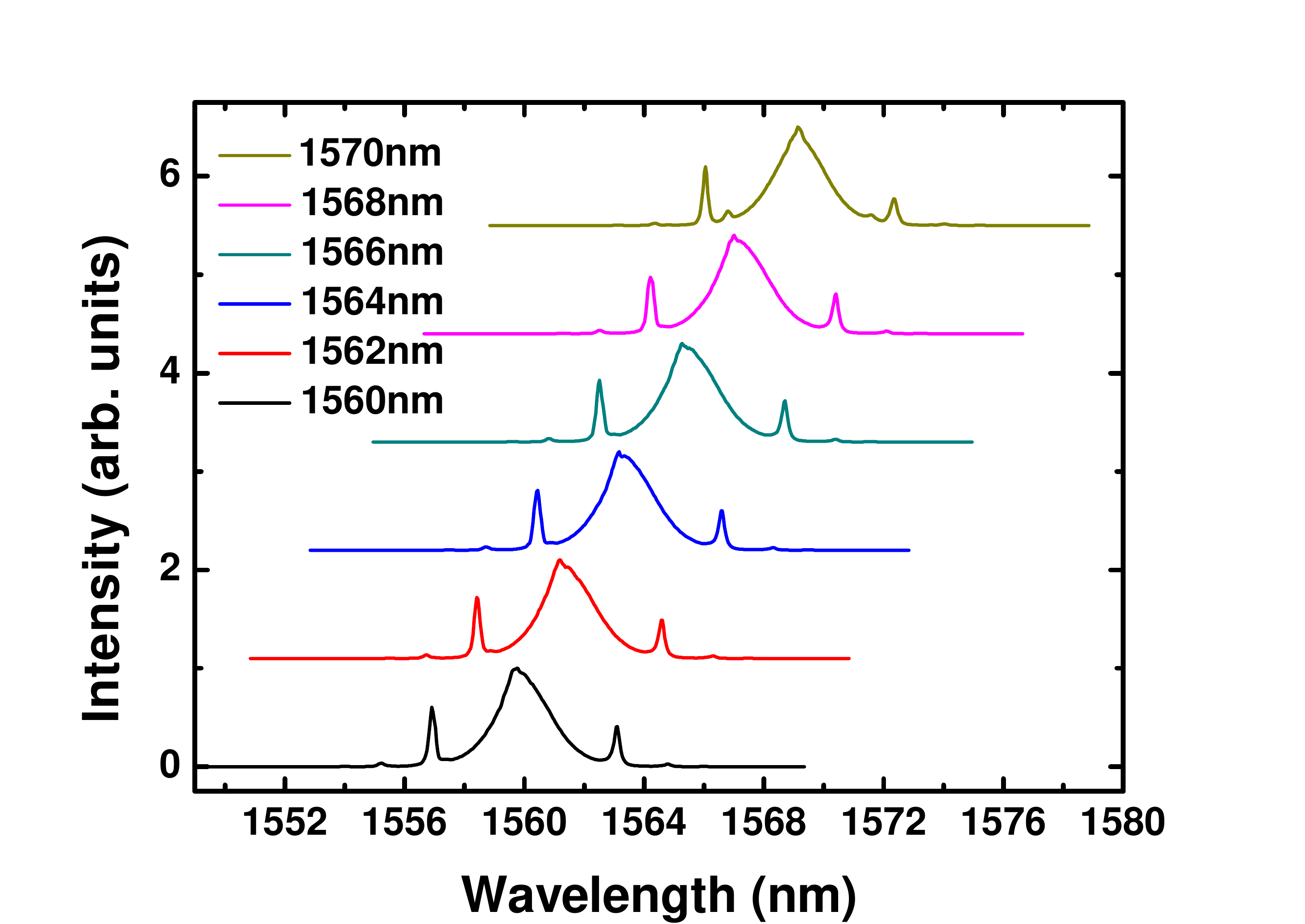}}
  \hspace{1in}
  \subfigure[]{
    \label{fig:subfig:b} 
    \includegraphics[width=8cm]{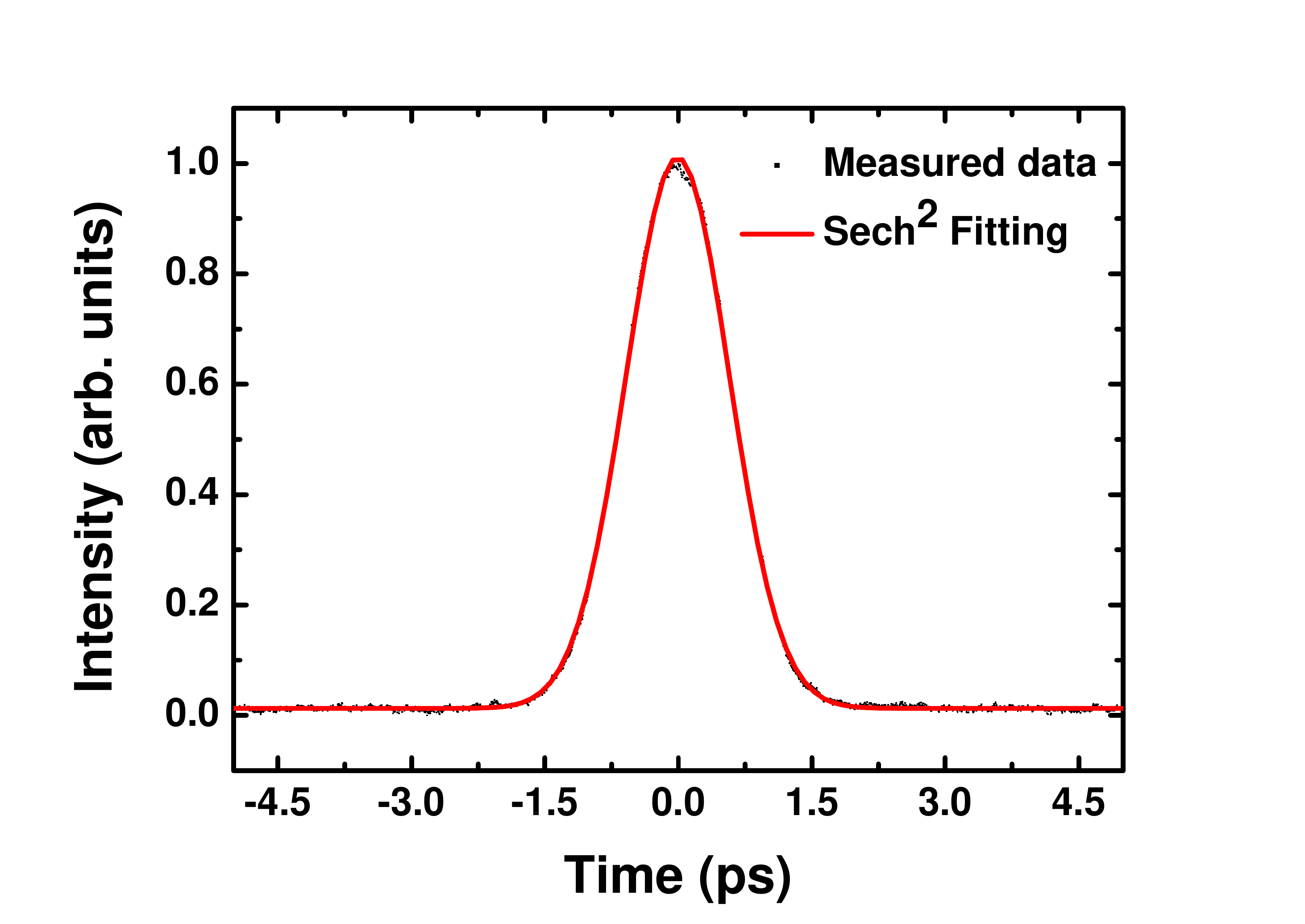}}
    \label{fig:subfig} 
  \caption{Mode locking performance in pure anomalous dispersion cavity. (a) Wideband output spectra from 1560 nm to 1570 nm and inserted numbers indicates the corresponding peak wavelengths. (b) Autocorrelation trace.}
\end{figure}

\subsection{Near zero dispersion cavity }

Operating in the near zero net cavity dispersion regime, the mode locking of the laser became difficult to achieve than in an anomalous dispersion fiber laser. It is compulsory to fine-tune the polarization controllers for starting the mode locking. However, once the laser is mode locked, the state can be maintained for long time. Fig. 3a shows the optical spectra and their wavelength tuning as the orientation of the intracavity PC was changed. The pulses have smooth spectrum, and as the orientation of the intra cavity PC was tuned, the central wavelength of the spectra could be continuously tuned from 1570 nm to 1600 nm.  The mode locked pulse spectra have a near-Gaussian profile, suggesting that they are the dispersion-managed solitons \cite{ref24}. The 3dB spectral bandwidth of the solitons varied under different central wavelengths, changed from $\sim$5.7 nm to $\sim$15.3 nm, and the corresponding soliton pulse width changed from $\sim$650 fs to $\sim$240 fs measured using a commercial autocorrelator (FR-103MN). Fig. 3b shows the measured autocorrelation trace of the narrowest pulses. It has a pulse with of $\sim$240 fs. The soliton spectrum is centered at $\sim$1598 nm, and the corresponding 3 dB spectral bandwidth is $\sim$15 nm, which gives a time-bandwidth product about 0.45.  The result indicates that the solitons are close to transform-limited.

\begin{figure}[htbp]
  \centering
  \subfigure[]{
    \label{fig:subfig:a} 
    \includegraphics[width=8cm]{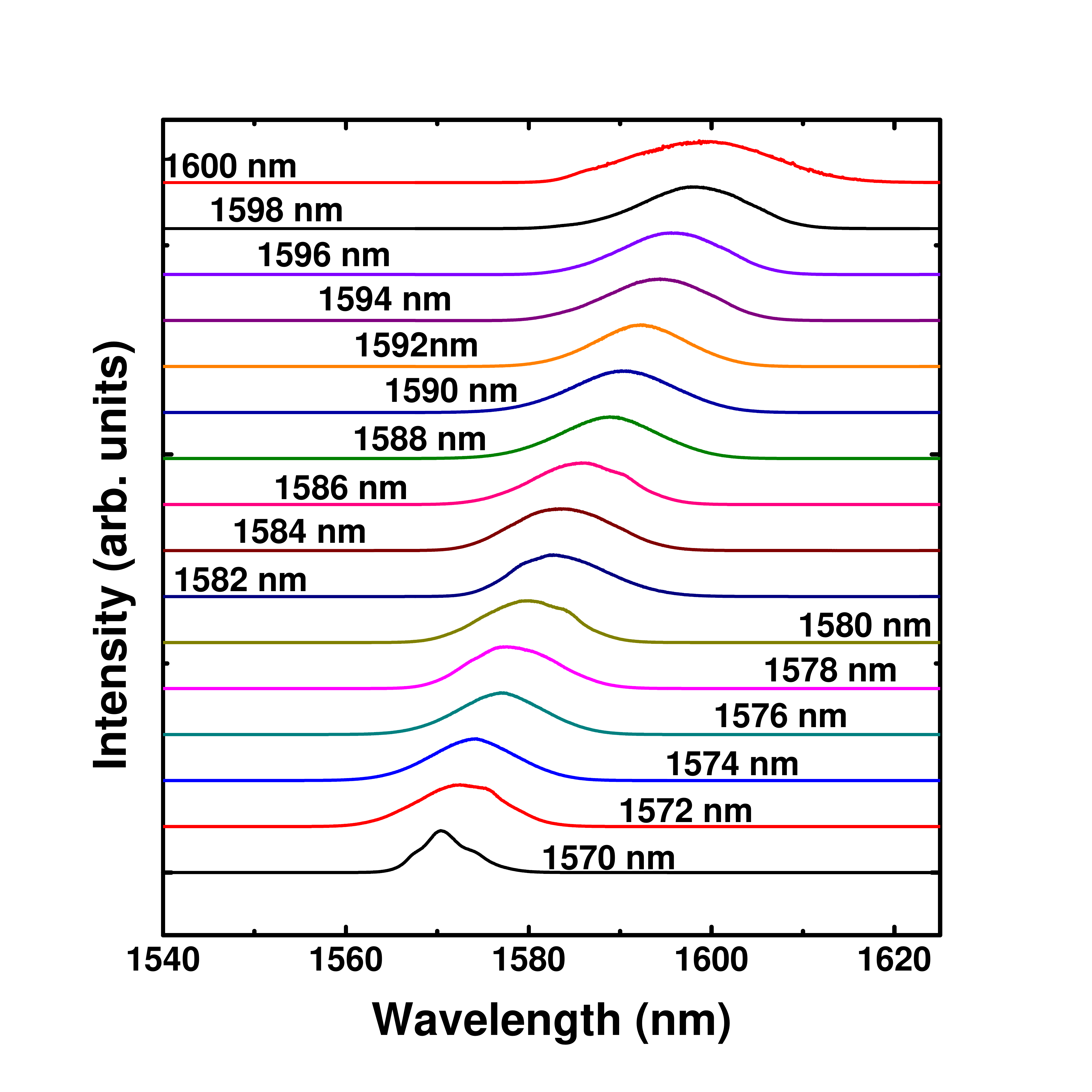}}
  \hspace{1in}
  \subfigure[]{
    \label{fig:subfig:b} 
    \includegraphics[width=8cm]{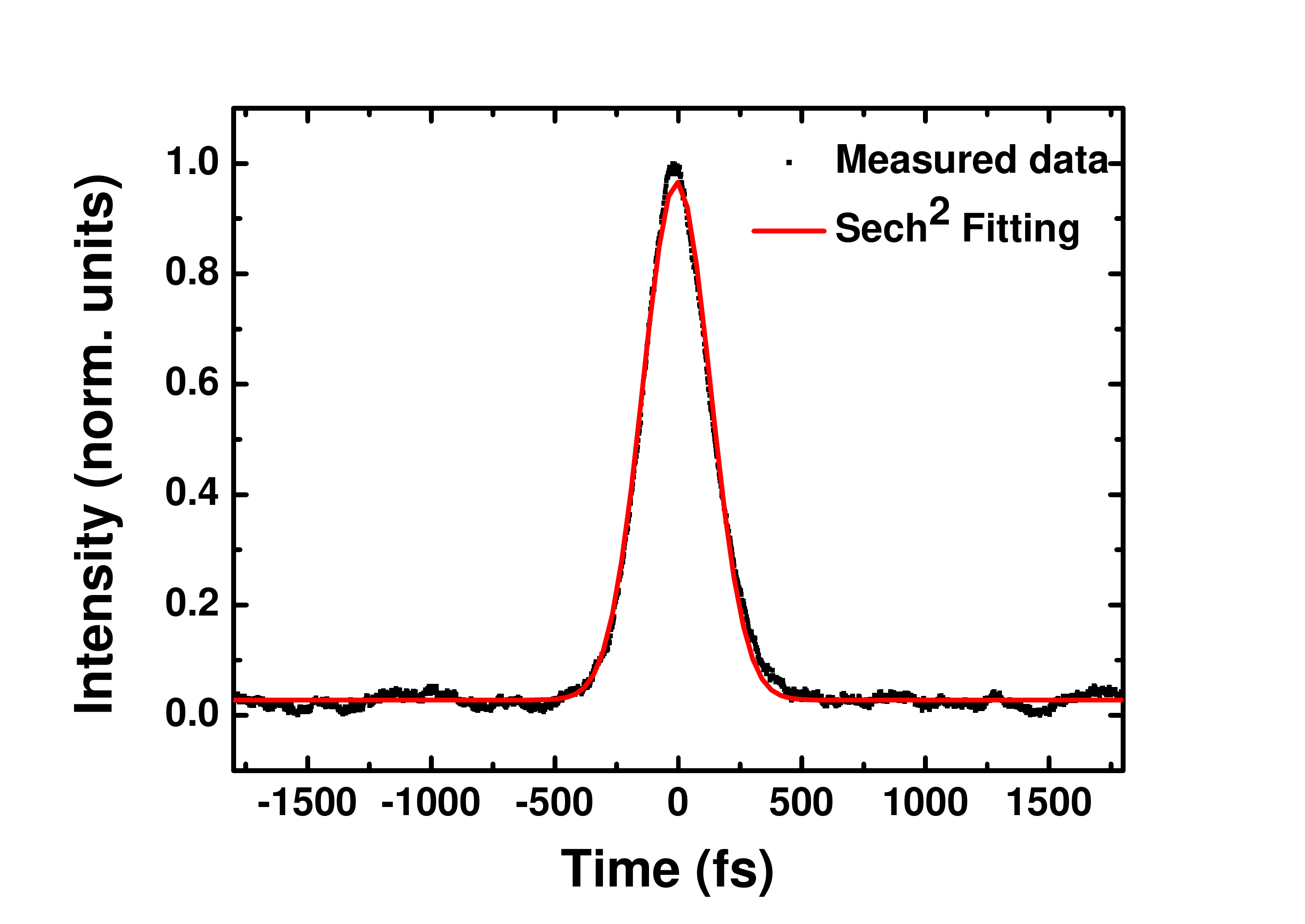}}
    \label{fig:subfig} 
  \caption{Mode locking performance in net-zero dispersion cavity. (a) Wideband output spectra from 1570 nm to 1600 nm. (b) Autocorrelation trace of the narrowest output.}
\end{figure}

\subsection{All normal dispersion cavity }

Finally, we operated the laser in the all-normal dispersion cavity regime. In the all-normal dispersion regime, dissipative solitons characterized by the characteristic steep spectral edges are always obtained. Fig. 4a shows the soliton spectra and the wavelength tuning of the dissipative solitons observed. Continuous wavelength-shift as large as $\sim$30 nm has been obtained. The 3 dB spectral bandwidth of the solitons varied with the central wavelength, changed from $\sim$3.0 nm to $\sim$6.0 nm, and the corresponding soliton pulse width changed from $\sim$ 150 ps to $\sim$70 ps, which was measured with a 50 GHz high-speed oscilloscope (Tektronix CSA 8000) together with a 45 GHz photo-detector (New Focus 1014). Fig. 4b shows a measured high-speed oscilloscope trace. It has a pulse width $\sim$150 ps, its central wavelength is at $\sim$1572 nm, and the 3 dB spectral bandwidth is $\sim$3.1 nm. The time-bandwidth product of the pulses is $\sim$58.1, indicating that the pulses are strongly chirped.

\begin{figure}[htbp]
  \centering
  \subfigure[]{
    \label{fig:subfig:a} 
    \includegraphics[width=8cm]{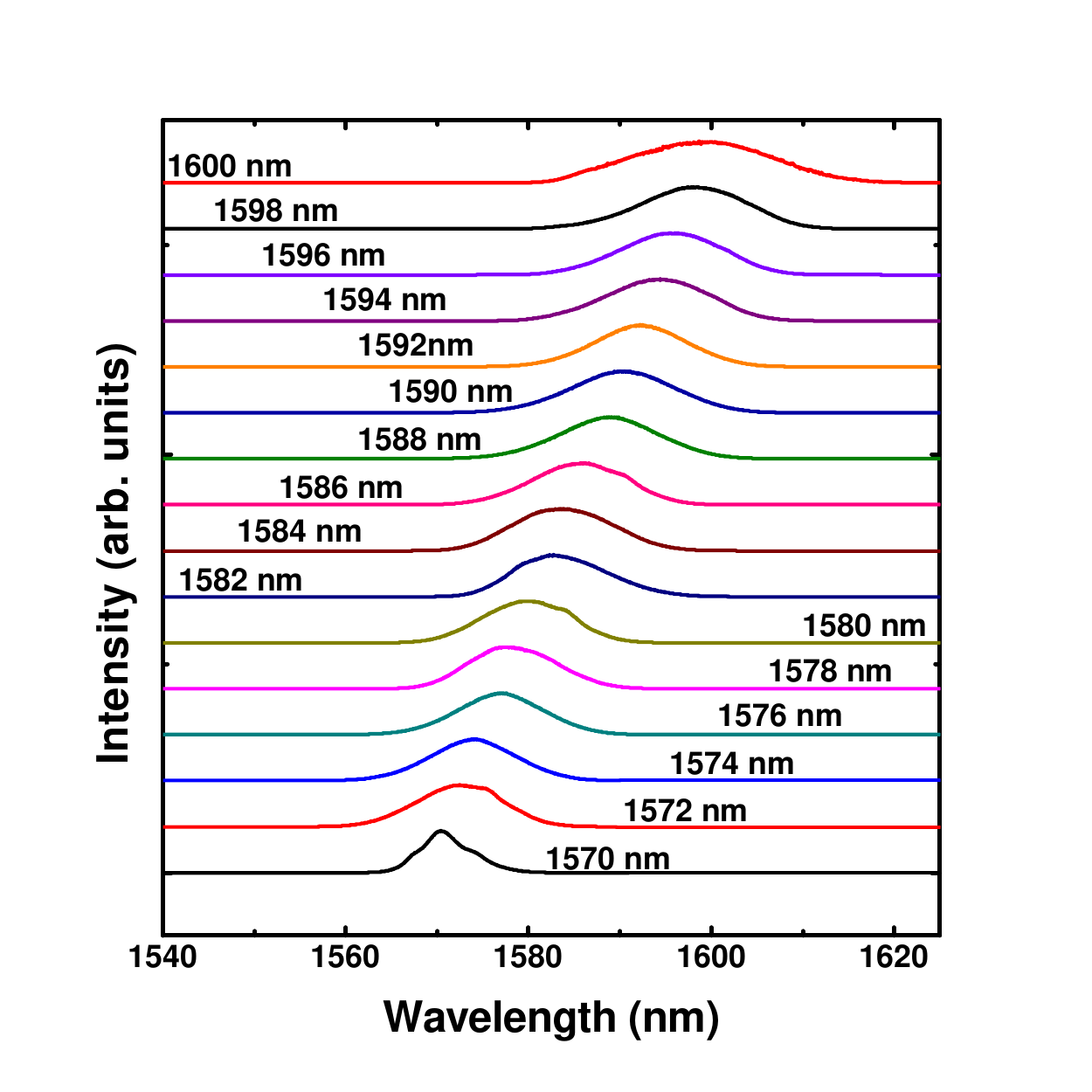}}
  \hspace{1in}
  \subfigure[]{
    \label{fig:subfig:b} 
    \includegraphics[width=8cm]{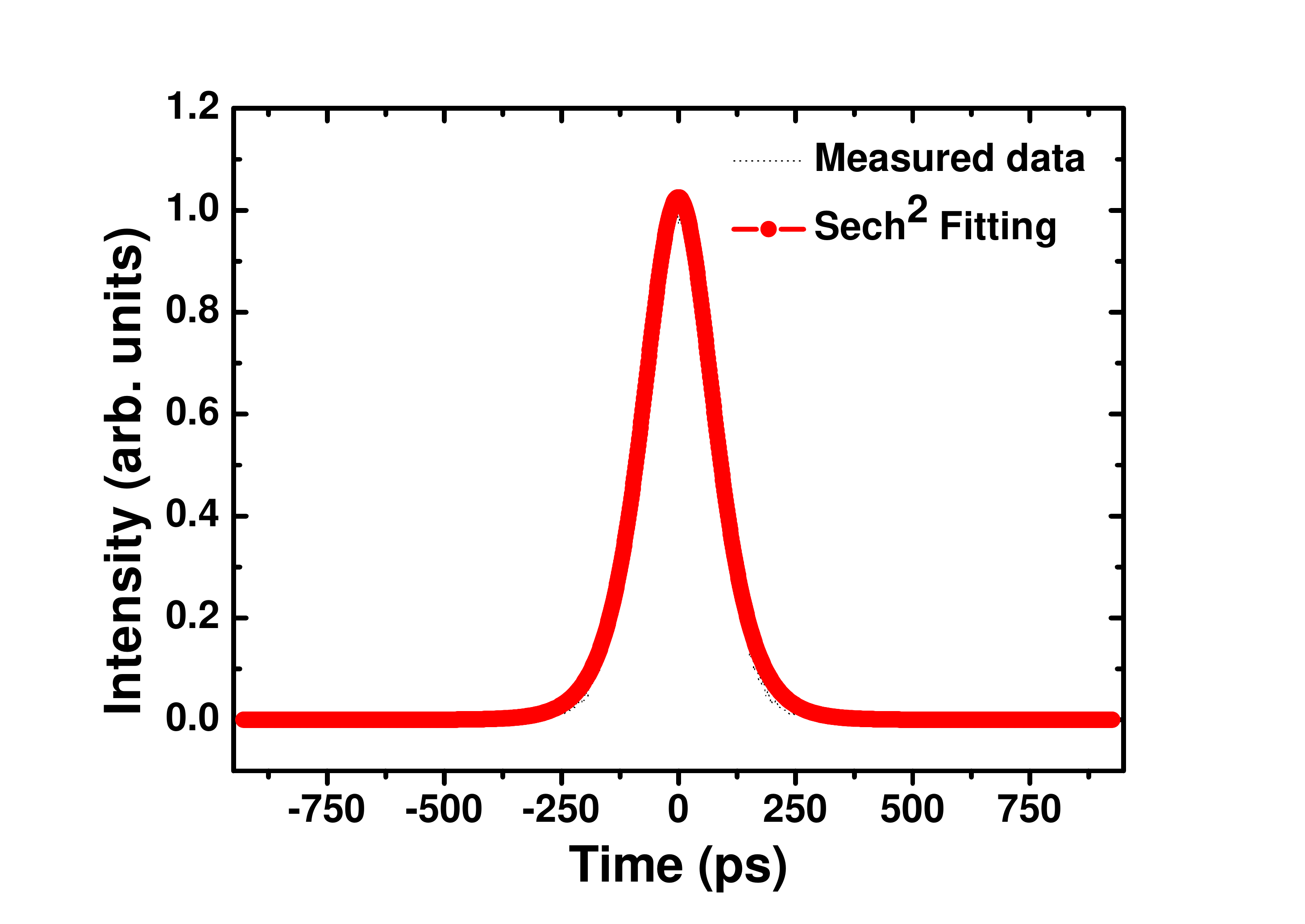}}
    \label{fig:subfig} 
  \caption{Mode locking performance in normal dispersion cavity. (a) Wideband output spectra from 1570 nm to 1600 nm. (b) Autocorrelation trace of the broadest output.}
\end{figure}

\section{Discussion}
Independent of the cavity dispersion, our experimental results have shown that under the graphene mode locking, continuous soliton central wavelength tuning of the laser emission could be obtained. A maximum soliton wavelength shift as large as $\sim$30 nm has been achieved in our experiments and the value of the maximum wavelength shift is limited by the effective gain bandwidth of the laser. In addition of the soliton central wavelength tuning, our experimental result have also shown that depending on the laser cavity design and its dispersion, the obtained stable soliton pulse width can also be varied in a wide range, from $\sim$240 fs to $\sim$150 ps in our experiments. The result demonstrates from another aspect the robustness of the graphene mode locked fiber lasers in terms of flexible optical pulse generation. We note that wavelength-tunable fs-to-ps, transform-limited to large frequency chirped pulses have widespread applications in many fields. The demonstrated graphene mode locked soliton fiber laser systems provide a user-friendly and cost effective solution for such a light source.

\section{Conclusion}
    In summary, graphene is a remarkable nano-material with broadband, controllable saturable absorption. Taking advantage of the broadband saturable absorption of graphene and the artificial cavity birefringence filter effect of fiber ring cavity, we have constructed wide range wavelength tunable erbium-doped soliton fiber lasers. We have shown that operating at different cavity dispersion regimes, not only the soliton wavelength could be tuned in a wider range, but also the soltion pulse width could be varied from the picosecond level to the femtosecons level. Our experimental results show that the graphene mode locked erbium-doped fiber lasers provide an easy way to generate stable ultrashort optical pulses with the desired pulse properties.
\\

\emph{Acknowledgments} Authors are indebted to Professor R. J. Knize from the United States Air Force Academy for useful discussions on graphene. The work is funded by the National Research Foundation of Singapore under the Contract No. NRF-G-CRP 2007-01. K. P. Loh wishes to acknowledge funding support from NRF-CRP Graphene Related Materials and Devices (R-143-000-360-281).

\newcommand{\ROM}[1]{\mathrm{\uppercase\expandafter{\romannumeral#1}}}


\begin{thebibliography}{10}
\bibitem{ref1} A. K. Geim, and K. S. Novoselov, Nat. Mater. \textbf{6}, 183-191 (2007).
\bibitem{ref2} Y.-M. Lin, C. Dimitrakopoulos, K. A. Jenkins, D. B. Farmer, H.-Y. Chiu, A. Grill, and Ph. Avouris, Science 327, \textbf{62} (2010).
\bibitem{ref3} R. W. Newson, J. Dean, B. Schmidt and H. M. van Driel, Opt. Express \textbf{17},  2326-2333  (2009).
\bibitem{ref4} F. N. Xia, T. Mueller, Y.M. Lin, A. V. Garcia and P. Avouris, Nat. Nanotechnol. \textbf{4}, 839-843 (2009).
\bibitem{ref5} Q. L. Bao, H. Zhang, Y. Wang, Z. Ni, Y. Yan, Z. X. Shen, K. P. Loh, D. Y. Tang, Adv. Funct. Mater. \textbf{19}, 3077- 3083 (2009).
\bibitem{ref6} Q. L. Bao, H. Zhang, J. X. Yang, S. Wang, D. Y. Tang, R. Jose, S. Ramakrishna, C. T. Lim and K. P. Loh, Adv. Funct. Mater. \textbf{20}, 782-791 (2010).
\bibitem{ref7} H. Zhang, Q. L. Bao, D. Y. Tang, L. M. Zhao, K. P. Loh, Appl. Phys. Lett. \textbf{95},141103 (2009).
\bibitem{ref8} H. Zhang, D. Y. Tang, L. M. Zhao, Q. L. Bao, K. P. Loh, Opt. Express \textbf{17},17630-17635 (2009).
\bibitem{ref9} W. D. Tan, C. Y. Su, R. J. Knize,  G.Q. Xie, L.J. Li and D. Y. Tang, Appl. Phys. Lett. \textbf{96}, 031106 (2010).
\bibitem{ref13} H. Zhang, D. Y. Tang, L. M. Zhao, X. Wu, Opt. Express \textbf{17}, 12692-12697 (2009).
\bibitem{ref10} N. N. Akhmediev, J. M. Soto-Crespo and Ph. Grelu, Phys. Lett. A \textbf{372}, 3124-3128 (2008).
\bibitem{ref11} L. E. Nelson, D. J. Jones, K. Tamura, H. A. Haus, and E. P. Ippen, Appl. Phys. B \textbf{65}, 277-294 (1997).
\bibitem{ref12} F. Wang, A. G. Rozhin, V. Scardaci, Z. Sun, F. Hennrich, I. H. White, W. I. Milne, and A. C. Ferrari, Nat. Nanotechnol. \textbf{3}, 738-742 (2008).
\bibitem{ref14} D. Y. Tang, H. Zhang, L. M. Zhao and X. Wu, Phys. Rev. Lett. \textbf{101}, 153904-153907 (2008).
\bibitem{ref15} Z. C. Luo, A. P. Luo, W. C. Xu, C. X. Song, Y. X. Gao, and W. C. Chen, Laser Phys. Lett. \textbf{6}, 582-585 (2009).
\bibitem{ref16} W. C. Chen, Z. C. Luo, and W. C. Xu, Laser Phys. Lett. \textbf{6}, 816-820 (2009).
\bibitem{ref17} L. R. Wang, X. M. Liu, and Y. K. Gong, Laser Phys.  Lett.\textbf{7}, 63-67 (2010).
\bibitem{ref18} X. M. Liu, Opt. Express, \textbf{17}, 22401 (2009); Phys. Rev. A \textbf{81},023811 (2010).
\bibitem{ref19} H. Zhang, D. Y. Tang, L. M. Zhao ,and N. Xiang, Opt. Express \textbf{16}, 12618-12623 (2008).
\bibitem{ref25} M. Zhang, L. L. Chen, C. Zhou, Y. Cai, L. Ren ,and Z. G. Zhang, Laser Phys. Lett. \textbf{6}, 657-660 (2009).
\bibitem{ref26} F. W. Wise, A. Chong, and W. Renninger, Laser Photonics Rev. \textbf{2}, 58-73 (2008).
\bibitem{ref27} X. M. Liu, T. Wang, C. Shu, L. R. Wang, A. Lin, K. Q. Lu, T. Y. Zhang and W. Zhao, Laser Physics \textbf{18}, 1357-1361 (2008).
\bibitem{ref21} H. Zhang, D. Y. Tang, L. M. Zhao, X. Wu, and H. Y. Tam, Opt. Express \textbf{17}, 455-460 (2009).
\bibitem{ref20} N. Akhmediev, and A. Ankiewicz, eds., Dissipative Solitons: From optics to biology and medicine, Lecture Notes in Physics, (Springer, Berlin-Heidelberg, 2008).
\bibitem{ref24} Y. S. Kivshar and G. P. Agrawal, Optical Solitons: From Fibers to Photonic Crystals (Academic, San Diego, 2003).



\end{thebibliography}
\end{document}